# Deterministic Bell state measurement with a single quantum memory

Akira Kamimaki[1,3], Keidai Wakamatsu[2], Kosuke Mikata[2], Yuhei Sekiguchi[1,3], and Hideo Kosaka[1,2,3]*

[1]Institute for Advanced Sciences (IAS), Yokohama National University, 79-5 Tokiwadai, Hodogaya, Yokohama 240-8501, Japan.

[2]Department of Physics, Graduate School of Engineering Science, Yokohama National University, 79-5 Tokiwadai, Hodogaya, Yokohama 240-8501, Japan.

[3]Quantum Information Research Center, Yokohama National University, 79-5 Tokiwadai, Hodogaya, Yokohama 240-8501, Japan.

*kosaka-hideo-yp@ynu.ac.jp

**Abstract**

**Any quantum information system operates with entanglement as a resource, which should be deterministically generated by a joint measurement known as complete Bell state measurement (BSM). The determinism arises from a quantum nondemolition measurement of two coupled qubits with the help of readout ancilla, which inevitably requires extra physical qubits. We here demonstrate a deterministic and complete BSM with only a nitrogen atom in a nitrogen-vacancy (NV) center in diamond as a quantum memory without reliance on any carbon isotopes by exploiting electron–nitrogen ($^{14}$N) double qutrits at a zero magnetic field. The degenerate logical qubits within the subspace of qutrits on the electron and nitrogen spins are holonomically controlled by arbitrarily polarized microwave and radiofrequency pulses via zero-field-split states as the ancilla, enabling the complete BSM deterministically. Since the system works under an isotope-free and field-free environment, the demonstration paves the way for realizing high-yield, high-fidelity, and high-speed quantum repeaters for long-haul quantum networks and quantum interfaces for large-scale distributed quantum computers.**

**Main Text**

The development of large-scale distributed quantum computers requires quantum networks[1-3] based on remote entanglement to connect the computers[4-10] and thus requires quantum repeaters[11-14] or quantum interfaces[15] that can perform a deterministic and complete Bell state measurement (BSM)[16-19] not only to extend the distance of photon transmission and to route photons over the networks but also to interface the quantum state between photons and qubits in quantum computers[15, 20-22]. A complete BSM allows us to project any two-qubit states into one of the four Bell states deterministically, which typically requires quantum nondemolition measurement known as single-shot measurement[23-28]. Due to quantum manipulability with communicating photons[29-34], as well as the coherence time of solid-state spins[33, 35-41], which is over a minute for a nuclear spin[41], nitrogen-vacancy (NV) centers in diamond[42-44] are of interest as core devices for quantum repeaters with quantum memories (Fig. 1a). The negatively charged NV center, in particular, has electron and nitrogen ($^{14}$N) nuclear spin composite systems, each forming a spin-1 triplet system that behaves as a three-level qutrit. An NV center also accompanies numerous carbon isotopes with a nuclear spin, forming a spin-1/2 doublet system that behaves as a two-level qubit. The measurement-based entanglement combined with a single-shot measurement was previously demonstrated by utilizing nitrogen nuclear spins and the nearby carbon isotope spins as the Bell states[16]. Subsequently, unconditional quantum teleportation between distant NV centers has been demonstrated based on the deterministic BSM[17], where the Bell states were composed of electron and nitrogen nuclear spins. The demonstrated Bell states with various combinations of spins were nondegenerate qubits energetically split by an external magnetic field more than 100 times stronger than the geomagnetic field and were dynamically controlled with resonant microwave (MW) and radiofrequency (RF) pulses.

However, the BSM at a zero magnetic field is desired for the quantum repeaters as an interface between communicating photons[45] and superconducting qubits[46-48], whose operation is restricted to a magnetic field sufficiently below superconducting breakdown for the sake of stability and uniformity. At a zero magnetic field, the electron and nitrogen-14 nuclear spins provide V-and Λ-shaped three-level qutrits with degenerate $m_s = \pm 1$ qubits and an energy-split $m_s = 0$ ancillary state component owing to the zero-field splitting of $D_0/2\pi = 2.88$ GHz for the electron (Fig. 1b) and the nuclear quadrupole splitting of $Q/2\pi = 4.95$ MHz for the nitrogen, respectively (Fig. 1c). The coupled-system Hamiltonian is given as

$$H = D_0 S_z^2 - Q I_z^2 - A S_z I_z, \qquad (1)$$

where $S_z$ and $I_z$ are the z component of the spin-1 operator of the electron spin and the nitrogen nuclear spin, respectively, and $A/2\pi = 2.17$ MHz is the hyperfine coupling between the two spins. The energy level diagram of the system is schematically shown in Fig. 1d. The Bell states are composed of inherently degenerate qubits, which we call geometric spin qubits[19, 28, 32, 34, 49-56], according to the computational basis states $|\pm 1\rangle_e$ for the electron and $|\pm 1\rangle_N$ for the nitrogen (dashed area in Fig. 1d), and are operated by the universal holonomic quantum gate with polarized MW and RF pulses via the ancilla states $|0\rangle_e$ and $|0\rangle_N$[55]. The geometric spin qubits $|\pm 1\rangle_e$ on an electron and $|\pm 1\rangle_N$ on a nitrogen atom are analogous to the polarization qubits on photons showing a geometric nature and have been demonstrated in terms of geometrically entangled emission[32] and absorption[34] of a photon. In this paper, we propose and experimentally demonstrate a novel scheme for the deterministic and complete BSM at a zero magnetic field with only a single quantum memory on a nitrogen atom in an NV center in diamond without relying on any carbon isotopes by exploiting the electron–nitrogen double qutrits at a zero magnetic field.

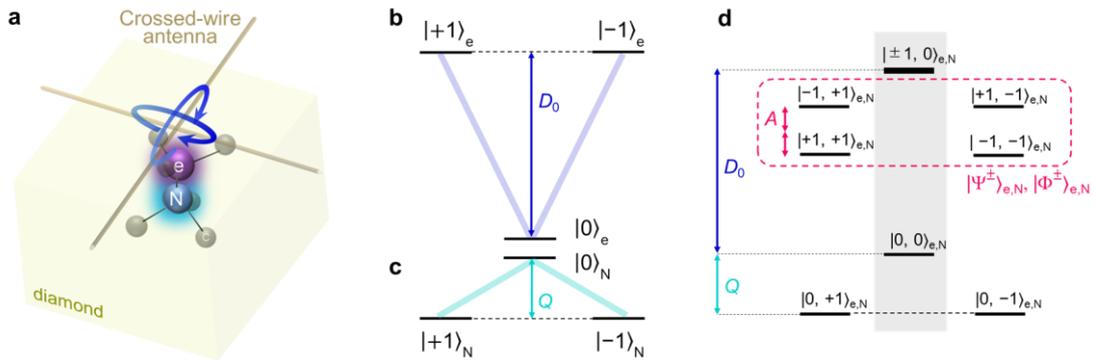

Figure. 1 | The concept of the scheme. Schematic of **a**, the diamond device, where electron and nitrogen ($^{14}$N) nuclear spins are manipulated by a crossed-wire antenna[55]. The energy diagram of the V-and Λ-shaped three-level qutrit systems

of the individual **b**, electron (e) and **c**, nitrogen nuclear (N) spins at a zero magnetic field. $D_0$ and $Q$ are the zero-field splitting of the electron and the nuclear quadrupole splitting of the nitrogen, respectively. **d**, The energy diagram of the double-qutrit joint states coupled with hyperfine interaction $A$. The four Bell states $|\Psi^\pm\rangle_{e,N}$ and $|\Phi^\pm\rangle_{e,N}$ on the logical qubits are based on the $|\pm 1, \pm 1\rangle_{e,N}$ computational bases (dashed area) and the $|0,0\rangle_{e,N}$ and $|\pm 1,0\rangle_{e,N}$ states serve as readout ancilla (shadowed area).

We use a single naturally occurring NV center in a high-purity type IIa chemical vapor deposition-grown diamond with a <100> crystalline orientation produced by Element Six. All measurements are performed at a temperature of under 5 K to allow coherent control of the electron orbital, and the sample is placed under an applied magnetic field with three-dimensional coils to suppress the geomagnetic field. Phonon sideband photons from an NV center are emitted and detected by the following optical setup. The system consists of a confocal microscope similar to those used in previous studies; it has a green-laser path (515 nm in wavelength) for non-resonant excitation and initialization of the charge state of an NV center and the electron spin as well as two red-laser paths (637 nm in wavelength) for the resonant excitation, initialization, and readout of the electron spin. In addition, the path for detecting the emitted photons is filtered by a dichroic mirror to exclude the green and red lasers and is focused on the avalanche photodiode (APD) to allow selective detection of phonon sideband photons.

The NV electron and nitrogen nuclear spins are individually manipulated with arbitrarily polarized MW and RF pulses created by two orthogonal wires[55] as shown in Fig. 1a. Here, the MW pulses are generated with the GRadient Ascent Pulse Engineering (GRAPE) algorithms, which enable high-fidelity operations of the geometric spin qubits[55, 57]. The MW (RF) pulses drive the unitary operations on the electron (nitrogen nuclear) spins in the Rabi frequency of the order of MHz (kHz). Figure 2a illustrates the quantum circuit for the complete BSM consisting of a disentanglement, which transforms the four Bell states into four eigenstates, and four sequential measurements of the four eigenstates. The Bell states defined as

$$|\Psi^{\pm}\rangle_{e,N} = \frac{1}{\sqrt{2}}(|+1,-1\rangle_{e,N} \pm |-1,+1\rangle_{e,N}),$$
$$|\Phi^{\pm}\rangle_{e,N} = \frac{1}{\sqrt{2}}(|+1,+1\rangle_{e,N} \pm |-1,-1\rangle_{e,N}),$$
(2)

are transformed as $|\Phi^+\rangle_{e,N} \rightarrow |+1,+\rangle_{e,N} \rightarrow |+1,+1\rangle_{e,N}$, $|\Phi^-\rangle_{e,N} \rightarrow |+1,-\rangle_{e,N} \rightarrow |+1,-1\rangle_{e,N}$, $|\Psi^+\rangle_{e,N} \rightarrow |-1,+\rangle_{e,N} \rightarrow |-1,+1\rangle_{e,N}$, and $|\Psi^-\rangle_{e,N} \rightarrow |-1,-\rangle_{e,N} \rightarrow |-1,-1\rangle_{e,N}$ by applying a MW pulse for the Controlled NOT (CNOT) gate and RF pulses for the Hadamard gate as depicted in Fig. 2b, where $|\pm\rangle_{e(N)} = \frac{1}{\sqrt{2}}(|+1\rangle_{e(N)} \pm |-1\rangle_{e(N)})$. It should be noted that the direct transition between $|\pm 1\rangle_{e(N)}$ states is not permitted, so the geometric nature is also utilized for the realization of the Hadamard gate[55]. Moreover, since the Hadamard gate for the nitrogen qubit uses a geometric phase in $|0\rangle_e$ subspace induced by the RF pulse[55], the electron qubit states $|\pm 1\rangle_e$ are sequentially transferred to $|0\rangle_e$ for applying the geometric phase and back to the original states. Finally, each of the resulting eigenstates (computational bases) $|\pm 1, \pm 1\rangle_{e,N}$ can be measured with quantum nondemolition readout by using an extra subspace in the three level systems (Figs. 2b,c,d). The details in the measurement are discussed later.

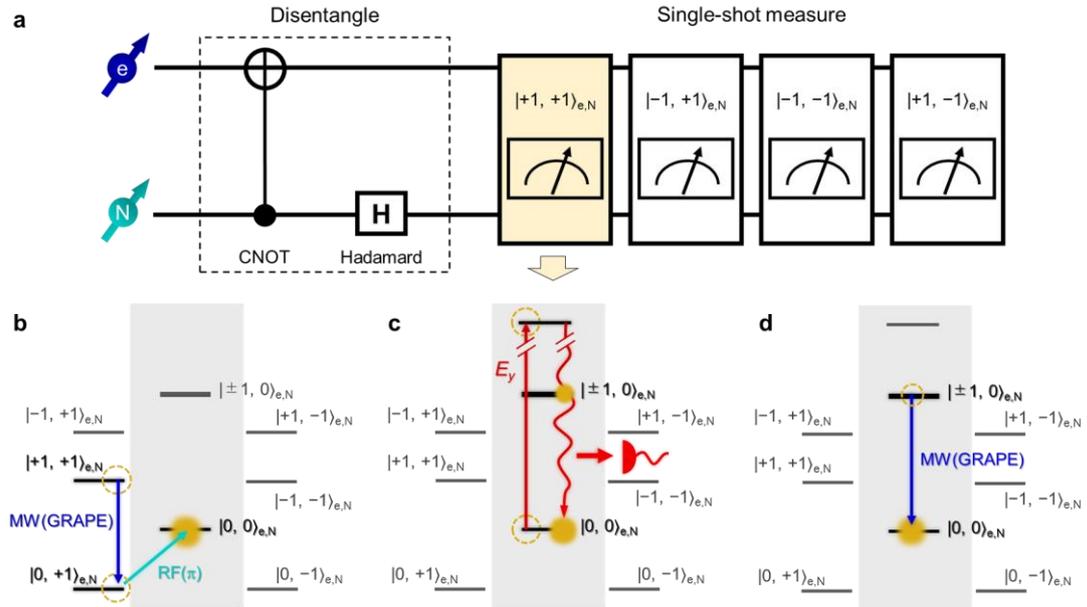

Figure. 2 | Scheme of the complete Bell state measurement (BSM). **a**, The quantum circuit utilizing the two-qubit joint states. After the entanglement generation, the BSM is achieved via the disentanglement operation consisting of CNOT (Controlled NOT) and Hadamard gate followed by a single-shot measurement. **b,c,d**, The procedure for measuring the $|+1,+1\rangle_{e,N}$ state. First, **b** each computational basis is selectively transformed to the $|0,0\rangle_{e,N}$

state by applying GRAPE (GRadient Ascent Pulse Engineering algorithms)-optimized microwave (MW) and radiofrequency (RF) pulses. Next, **c** the population is read out by irradiating a red laser pulse. Finally, **d** the probabilistic transition states of $|\pm 1, 0\rangle_{e,N}$ are initialized to the $|0,0\rangle_{e,N}$ state by irradiating a GRAPE-optimized MW pulse.

We initially evaluate the process of disentanglement by quantum state tomography (QST) of the four prepared Bell states, the states after the CNOT gate, and the states after the Hadamard gate. As shown in Fig. 3a, QST consists of a transfer of the arbitrary state $|\psi_e, \psi_N\rangle_{e,N}$ into the $|0,0\rangle_{e,N}$ state and the repetitive readout of the nuclear spin state via electron spin. Initially, the GRAPE-optimized MW pulse transforms the $|\psi_e\rangle_e$ state, which is selected by the polarization of the MW, into the $|0\rangle_e$ state regardless of the nuclear spin state. Next, the RF pulse transforms the $|\psi_N\rangle_N$ state, which is also selected by the polarization of the RF, into the $|0\rangle_N$ state conditioned on the electron spin state of $|0\rangle_e$. As a result, the population of the target state is stored in $|0\rangle_N$ and the others remain in $|\pm 1\rangle_N$, allowing repetitive readout of the target state via the nuclear spin. The sub-sequence of the readout repeated 30 times consists of the initialization of the electron spin, the mapping of the nuclear spin state into the electron spin state, and the readout of the electron spin. The initialization is performed by spin pumping into the $|0\rangle_e$ state by $|\pm 1\rangle_e$-selective excitation to the $E_{1,2}$ excited state. The mapping is performed by the GRAPE-optimized MW pulse to flip the $|0\rangle_e$ state into the $|\pm 1\rangle_e$ state conditioned on the nuclear spin state of $|0\rangle_N$. The readout is performed by counting the photons of phonon sideband emission during $|0\rangle_e$-selective excitation to the $E_y$ excited state. Figs. 3b,c,d show the reconstructed density matrices of the $|\Psi^\pm\rangle_{e,N}$ and $|\Phi^\pm\rangle_{e,N}$ states after entanglement generation, $|\pm 1, \pm\rangle_{e,N}$ states after the CNOT gate operation, and $|\pm 1, \pm 1\rangle_{e,N}$ states after the Hadamard gate operation, respectively (fidelities are shown in the figures). The obtained fidelities exceed 90% on average for all three stages: the prepared Bell states, states after the CNOT gate, and states even after the Hadamard gate.

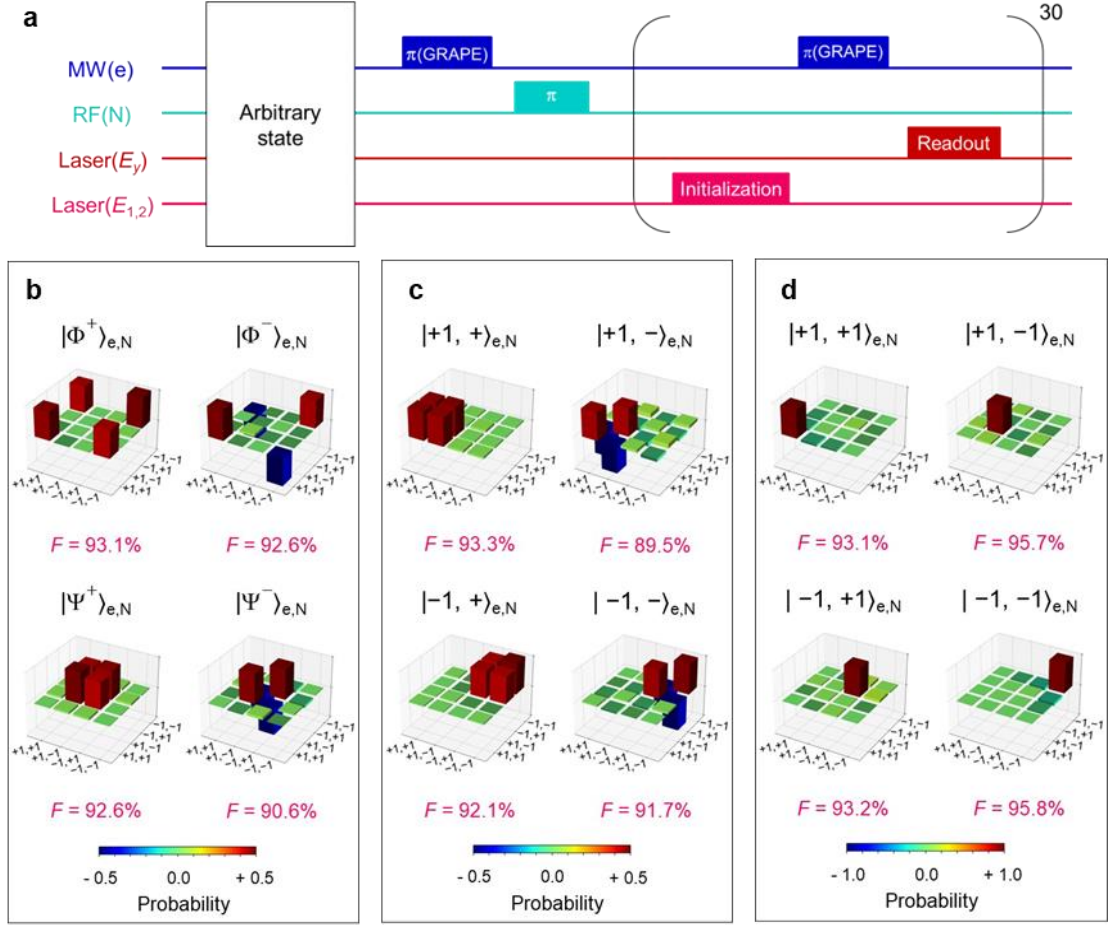

Figure. 3 | Quantum state tomography (QST). **a**, The pulse sequence of QST. An arbitrary state is mapped to the $|0\rangle_N$ state and repeatedly read out by $E_y$ resonant red laser (100 nW). The real part of the QST **b** after Bell state generation, **c** after CNOT gate and **d** after Hadamard gate. The basis labeled $\pm 1, \pm 1$ for each the state corresponds to the $|\pm 1, \pm 1\rangle_{e,N}$ state, respectively. Note that the RF-pulse polarizations for the final state (**d**) are optimized, slightly increasing the fidelity compared with the others (**b**, **c**). For all states, the obtained fidelities $F$ exceed 90%.

We now demonstrate the complete BSM, which enables deterministic discrimination of all four Bell states with only one set of measurements. Figure 4a illustrates the pulse sequence for the BSM. The measurements are carried out in the order of $|\Phi^+\rangle_{e,N}$, $|\Psi^+\rangle_{e,N}$, $|\Psi^-\rangle_{e,N}$, and $|\Phi^-\rangle_{e,N}$ by selecting the MW-pulse frequency and RF-pulse polarization (R or L) to transfer them into the corresponding computational bases $|\pm 1, \pm 1\rangle_{e,N}$, which are then selectively transformed again into the readout state $|0,0\rangle_{e,N}$ by the polarized MW and RF pulses. The $|0,0\rangle_{e,N}$ state is repeatedly measured in the ancillary space $\{|0,0\rangle_{e,N}, |\pm 1,0\rangle_{e,N}\}$. In contrast to the case of QST, initialization by $E_{1,2}$ excitation is no longer available since it disrupts the state awaiting the next measurement. Instead of using $E_{1,2}$ lasers, we use the GRAPE-optimized MW pulses to transform a part of the $|\pm 1, 0\rangle_{e,N}$ states relaxed by the $E_y$ excitation during the measurement of $|0,0\rangle_{e,N}$ (here, the $|+,0\rangle_{e,N}$ state) back into the $|0,0\rangle_{e,N}$, allowing for repetitive measurements of all the computational bases $|\pm 1, \pm 1\rangle_{e,N}$ through the readout state $|0,0\rangle_{e,N}$. Photon counts are accumulated by repeating the sub-sequences 25 times in a similar way as for the QST. The Bell states are discriminated by the conditions as

$$
\begin{aligned}
&|\Phi^+\rangle_{e,N} : n_1 \geq n_c, \\
&|\Psi^+\rangle_{e,N} : n_1 < n_c, n_2 \geq n_c, \\
&|\Psi^-\rangle_{e,N} : n_1 < n_c, n_2 < n_c, n_3 \geq n_c, \\
&|\Phi^-\rangle_{e,N} : n_1 < n_c, n_2 < n_c, n_3 < n_c, n_4 \geq n_c
\end{aligned}
\quad (3)
$$

where $\{n_1, n_2, n_3, n_4\}$ are the photon counts for the successive measurements in the $|\Phi^+\rangle_{e,N}$, $|\Psi^+\rangle_{e,N}$, $|\Psi^-\rangle_{e,N}$, and $|\Phi^-\rangle_{e,N}$ bases and $n_c$ is the threshold of the photon counts, which is set to $n_c = 1$ in this demonstration. Figure 4b shows the probability distributions of the accumulated photon counts. The distribution clearly changes from a dark state (0.3 on average) to a bright state (1.8 on average) when the prepared Bell state corresponds to the measurement state, indicating that the Bell states are well discriminated by Eq. (3). It should be noted that the distribution is kept bright for the following measurements after the correspondence since the population of the measured state remains in the readout ancilla. The final measurement in $|\Phi^-\rangle_{e,N}$ is confirmed to be bright as $n_4 \geq n_c$, although it should be determined by three measurements. All four Bell states are thus equivalently discriminated deterministically. Figure 4c shows the probability distributions of the measurement outcome after the thresholding discrimination. Note that the discriminated Bell states (red bars) correspond to the prepared Bell states with a fidelity of $F_{BSM} = 68\%$ on average.

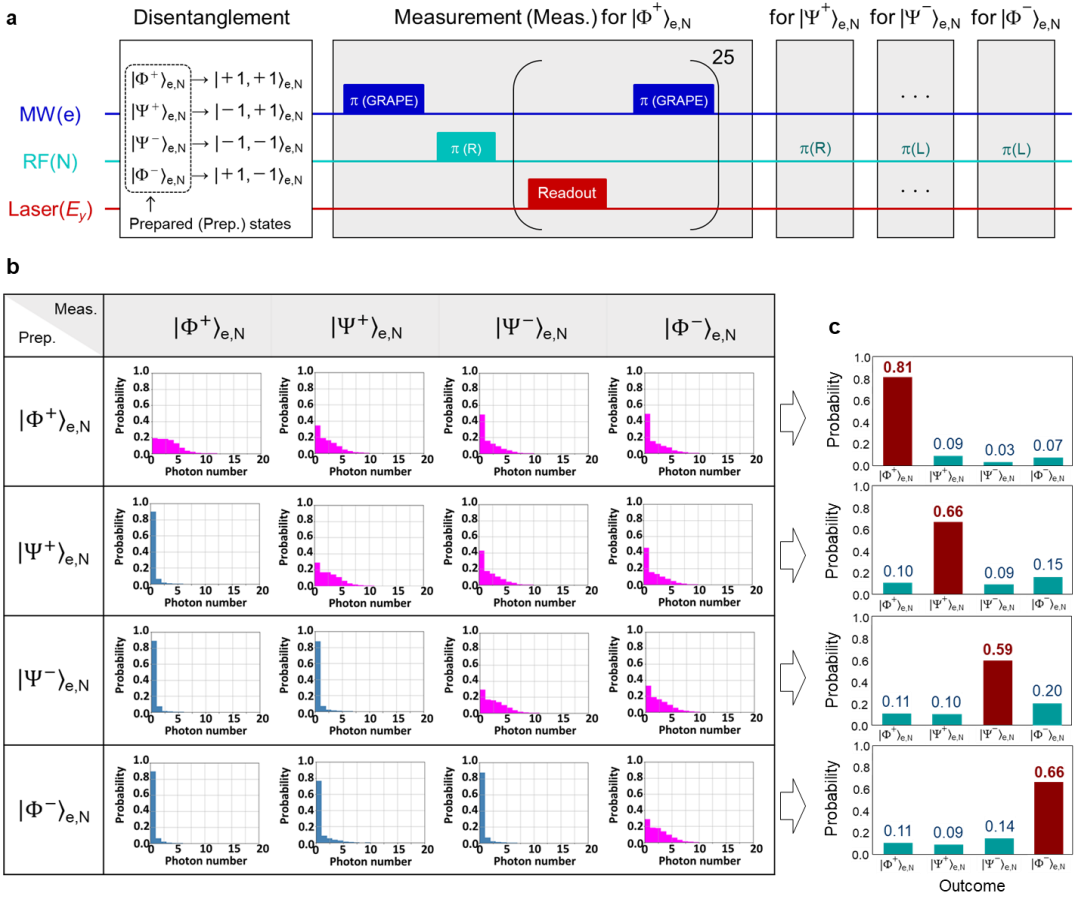

Figure. 4 | Complete BSM. **a**, The pulse sequences of the single-shot measurements ($|\Phi^+\rangle_{e,N}$ measurement is only described as an example). The BSM consists of selective transformation from the prepared Bell state into the readout state $|0,0\rangle_{e,N}$ through the computational basis $|\pm1,\pm1\rangle_{e,N}$ by corresponding GRAPE-shaped MW and polarized square-shaped RF π-pulses followed by a repetitive single-shot measurement by $E_y$ laser pulses for the measurement of $|0,0\rangle_{e,N}$ and GRAPE-based MW π-pulses for the initialization (instead of $E_{1,2}$ lasers), respectively. The pulse sequences for the other measurements are the same as those for $|\Phi^+\rangle_{e,N}$ (shown in the shadowed area) except for the GRAPE waveform (targeting upper or lower level) and the RF polarization (R or L) depending on the prepared Bell state. **b**, The probability distribution of the accumulated photon counts for the BSM. Magenta (blue) bars indicate that the obtained photon counts exceed (fall below) a threshold. **c**, Probability distributions of the Bell states after the thresholding discrimination. The red bars indicate that the discriminated state corresponds to the prepared Bell state.

Without transitioning to ancilla $|0\rangle_N$ for the readout, it is difficult to determine the prepared state using electron spins; this is due to the depolarization of the electron spin by the laser exposure and indicates the significance of the qutrit nature with the readout ancilla. Note that the small extraction efficiency of photons emitted from bulk diamond due to a high refractive index requires many repetitive measurements without a solid immersion lens (SIL), degrading the fidelity unless the RF, MW, and laser pulses are ideally optimized. Moreover, the low extraction efficiency leads to the deviation from the Poisson distribution even for the bright states exceeding the threshold photon counts as shown in Fig. 4b. The deviation should be improved by increasing the efficiency with the use of a SIL[17].

The demonstrated scheme plays a complementary role to the conventional scheme. The complete BSM of the electron and nitrogen nuclear spins in an NV center was previously demonstrated by W. Pfaff *et al.*, where the spin states of the $|0\rangle_{e(N)}$ and $|-1\rangle_{e(N)}$ are mainly utilized[17]. They achieved a BSM with a monolithic diamond SIL under magnetic fields, which enabled the single-shot measurement of the electron spin itself and reduced the repetition time of the single-shot measurement. In contrast, our scheme relies on neither a SIL nor magnetic fields. Application of magnetic fields prevents the integration of the quantum interface with superconductive qubits[46-48], which plays a key role in distributed quantum computers. Instead of relying on high photon extraction efficiency to directly measure electron spin, our scheme fully exploits the inherent qutrit nature of spins in an NV center serving as the Bell bases and the readout ancilla for the single-shot measurement.

In summary, a deterministic and complete BSM has been demonstrated at a zero magnetic field with only a single quantum memory by fully exploiting the inherent qutrit nature of electron and nitrogen spins in an NV center. The double qutrit systems enabled nondestructive joint-state measurements without relying on extra carbon isotopes or high photon extraction efficiency from an electron, owing to the long memory time of the nitrogen nuclear spin. The present demonstration paves the way for realizing high-yield, high-fidelity, and high-speed quantum repeaters for long-haul quantum networks and quantum interfaces for large-scale distributed quantum computers with minimal physical resources.

We thank H. Kato, T. Makino, T. Teraji, Y. Matsuzaki, K. Nemoto, N. Mizuochi, F. Jelezko and J. Wrachtrup for their discussions and experimental help. This work was supported by Japan Society for the Promotion of Science (JSPS) Grants-in-Aid for Scientific Research (grant numbers 20H05661 and 20K2044120); by the Japan Science and Technology Agency (JST) CREST (grant number JPMJCR1773); and by JST Moonshot R&D (grant number JPMJMS2062). We also acknowledge the assistance of the Ministry of Internal Affairs and Communications (MIC) under the initiative Research and Development for Construction of a Global Quantum Cryptography Network (grant number JPMI00316).